\documentclass[epj, final]{svjour}

\usepackage{palatino}
\usepackage[latin1]{inputenc}
\usepackage{epsf}
\usepackage{amsmath,amssymb}
\usepackage{latexsym}
\usepackage{calc}
\usepackage{color}
\usepackage{shadow}
\usepackage{epsfig}
\usepackage{float}
\usepackage{braket}

\usepackage{dsfont}

\usepackage[pdftex, pdfstartview = {FitH}]{hyperref}

\newcommand{\ben}{\begin{equation}}
\newcommand{\een}{\end{equation}}
\newcommand{\bea}{\begin{eqnarray}}
\newcommand{\eea}{\end{eqnarray}}

\def\br{{\bf r}}
\def\bR{{\bf R}}
\def\bP{{\bf P}}
\def\bA{{\bf A}}

\def\bp{{\bf p}}

\def\dulR{{\underline{\underline{\bf R}}}}
\def\dulr{{\underline{\underline{\bf r}}}}
\def\dulq{{\underline{\underline{\bf q}}}}

\def\orr{\overrightarrow}

\begin{document}
\title{Light-Matter Interactions via the Exact Factorization Approach}

\author{Norah M. Hoffmann\inst{1,2} \and Heiko Appel\inst{1} \and Angel Rubio\inst{1} \and Neepa T. Maitra\inst{2,3}}
\institute{Max Planck Institute for the Structure and Dynamics of Matter and
Center for Free-Electron Laser Science and Department of Physics,
Luruper Chaussee 149, 22761 Hamburg, Germany
\and Department of Physics and Astronomy, Hunter College of the City University of New York, 695 Park Avenue, New York, New York 10065, USA
\and The Physics Program and the Chemistry Program of the Graduate Center of the City University of New York, New York 10065, USA} 
%\email{nmaitra@hunter.cuny.edu; 
\date{\today}
%\pacs{31.15.-p, 31.50.-x, 82.20.Gk}

\abstract
{ The exact factorization approach, originally developed for electron-nuclear dynamics, is extended to light-matter interactions within the dipole approximation. This allows for a  Schr\"odinger equation for the photonic wavefunction, in which the potential contains exactly the effects on the photon field of its coupling to matter. We illustrate the formalism and potential for a two-level system representing the matter, coupled to an infinite number of photon modes in the Wigner-Weisskopf approximation, as well as a single mode with various coupling strengths. Significant  differences are found with the potential used in conventional approaches, especially for strong-couplings. We discuss how our exact factorization approach for light-matter interactions can be used as a guideline to develop semiclassical trajectory methods for efficient simulations of light-matter dynamics.
}

\maketitle

\section{Introduction}

The interaction of light with matter involves the correlated dynamics of photons, electrons, and nuclei. Even at a non-relativistic level the solution of Schr\"odinger's equation for the coupled subsystems is a daunting computation. In a given situation however, one is often measuring properties of only one of these subsystems. For example, one might be wanting to know how the electrical conductivity of a molecule is affected by the photons, as in the recent experiment showing the increased conductivity of organic semiconductors due to hybridization with the vacuum field~\cite{Orgiu15}.  On the other hand, one might want to understand how molecular dissociation after electronic excitation is affected in the presence of light, as in the recent study of light-induced versus intrinsic non-adiabatic dynamics in diatomics \cite{CHCV17}. Or, one might want to measure the superradiance from a collection of atoms \cite{GH82}. In each of these three cases, the observable of interest involves one of the subsystems alone, electronic, nuclear, and photonic, respectively, yet to capture the dynamics of the relevant subsystem, clearly the effects of all subsystems are needed.
The question then arises: can we write a Schr\"odinger equation for one of the subsystems alone, such that the solution yields the wavefunction of that subsystem? The potential appearing in the equation would have to incorporate the couplings to the other subsystems as well as to any externally applied fields. 

Hardy Gross, with co-workers, in fact already \linebreak answered exactly these questions~\cite{GG14,AMG10,AMG12} when the photonic system is treated as a {\it classical} light field neglecting the magnetic field contribution. That is, for systems of electrons and nuclei, interacting with each other via a scalar potential (usually taken as Coulomb), and in the presence of an externally applied scalar potential, such as the electric field of light, it was shown that one can exactly factorize the complete molecular wavefunction into a wavefunction describing the nuclear system, and a wavefunction describing the electronic system that is conditionally dependent on the nuclear subsystem \cite{GG14,AMG10,AMG12,H75}: $\Psi(\dulr,\dulR,t) = \chi(\dulr,t)\Phi_\dulR(\dulr,t)$, where $\dulr = \br_1,...\br_{N_e}$ and $\dulR = \bR_1...\bR_{N_n}$ represent all electronic and nuclear coordinates respectively. The equation for the nuclear subsystem has a Schr\"odinger form, with scalar and vector potentials that completely account for the coupling to the electronic system. One can reverse the roles of the electronic and nuclear subsystems, to instead get a Schr\"o- dinger equation for the electronic system, which is particularly useful when one is most interested in the electronic properties~\cite{SAMYG14}, e.g. in field-induced molecular ionization. 

Recently rapid experimental and theoretical advances have however drawn attention to fascinating phenomena that depend on the quantization of the light field in its interaction with matter. This includes few-photon coherent nonlinear optics with single molecules \cite{maser2016}, direct experimental sampling of electric-field vacuum fluctuations \cite{riek2015,moskalenko2015}, multiple Rabi splittings under ultrastrong vibrational coupling \cite{george2016}, exciton-polariton condensates \cite{byrnes2014,kasprzak2006}, polaritonically enhanced superconductivity in cavities \cite{SRR18}, or frustrated polaritons \cite{Schmidt2016} among others.
Optical cavities can be used to tune the effective strength of the light-matter interaction, and, in the strong-coupling regime in particular, one finds for example non-radiative energy transfer well beyond the F\"orster limit between spatially separated donors and acceptors \cite{zhong2017}, strong coupling between chlorosomes of photosynthetic bacteria and confined optical cavity modes \cite{coles2014}, photochemical reactions can be suppressed with cavity modes \cite{Galego2016}, the position of conical intersections can be shifted or they can be removed \cite{Flick2017,CHCV17}, or state-selective chemistry at room temperature can be achieved by strong vacuum-matter coupling \cite{ebbesen2016}.
Strong vacuum-coupling can change chemical reactions, such as photoisomerization or a prototypical deprotection reaction of alkynylsilane \cite{ebbesen2016,thomas2016}
This has given rise to the  burgeoning field now sometimes called "polaritonic chemistry" \cite{BGA18,HS18,FGG18,ribeiro2018polariton}. 
In addition, novel spectroscopies have been proposed which explicitly exploit correlated states of the photon field. For example 
the use of entangled photon pairs enables one to go beyond the classical Fourier limit \cite{dorfman14}, or correlated photons can be used to imprint
correlation onto matter\cite{Flick2017,Ficek1997,Matsukevich2004}

In this paper, we extend the exact factorization approach to  non-relativistic coupled photon-matter systems within the dipole approximation.  We focus particularly on finding the potential driving the system in the present study. One motivation is towards developing mixed \linebreak quantum-classical methods for the light-matter system. The observation that in a matter-free system, the photonic Hamiltonian is a sum over harmonic Hamiltonians for each mode of the radiation field suggests that a classical treatment of the photonic system would be accurate: if the system begins in a Gaussian wavepacket, classical Wigner dynamics exactly describes the motion \cite{Heller76}. Coupling to matter within the dipole approximation where the coupling operator is linear in the photonic variable preserves the quadratic nature of the Hamiltonian, and one might then think that again a classical Wigner treatment would be exact. However, recently it was found that, although accurate, it was not exact~\cite{Norah_inprep}. This implies that the true potential driving the photonic motion is in fact not quadratic. The exact factorization approach defines exactly what this potential should be. In this paper we explain the formalism and give some examples of this potential, that clearly show deviations from harmonic behaviour throughout the dynamics. 

The theory is described in Sec.~\ref{sec:theory}, presenting the Hamiltonian that we will consider, and the formalism of the factorization approach. Section~\ref{sec:results} demonstrates the approach on two examples, that we choose as the simplest cases for this initial study. The matter system is described by a two-level system while the photonic system is chosen to either be an infinite number of modes treated within the Wigner-Weisskopf approximation, or a single cavity mode chosen to be resonant with the spacing of the two levels, explored over a range of coupling strengths. We find and interpret the potential driving the photonic system, which depends significantly on whether the initial state of the system is chosen correlated or fully factorized. Finally in Sec.~\ref{sec:summary} we summarize and discuss the relevance of this approach for future investigations of light-matter dynamics. 
 
%\begin{center}
\section{Theory}
\label{sec:theory}
%\end{center}

\subsection{QED-Hamiltonian}
In this work, we consider the non-relativistic limit of a system of  $N_{e}$ electrons, $N_{n}$ nuclei, and $N_{p}$ quantized photon modes, treated within the dipole approximation in Coulomb gauge~\cite{Faisal1987,Flick2017a,ruggenthaler2018}. For now, we do not consider any classical external fields, and neglect spin-coupling. The Hamiltonian  of this coupled system is then defined by \cite{Tokatly2013,Flick2017,Pellegrini2015,Flick2015,Craig1998}
\begin{equation}
\hat{H}(\dulq,\dulr,\dulR) = \hat{H}_{p}+ \hat{H}_{e}  + \hat{H}_{n}+ \hat{H}_{ep} +  \hat{H}_{np} + \hat{H}_{en} +  \hat{H}_{pen},
\label{fullH}
\end{equation}
which operates in the space of:  $\dulr = \{\br_1..\br_i..\br_{N_e}\}$ \linebreak representing all electronic spatial coordinates, \linebreak $\dulR= \{\bR_1..\bR_I..\bR_{N_n}\}$ representing all nuclear coordinates, and $\dulq= \{q_1..q_\alpha..q_{N_p}\}$ representing all photonic displacement coordinates.
The first term characterizes the cavity-photon Hamiltonian
\begin{equation}
\hat{H}_{p}(\dulq) = \frac{1}{2}\left(\sum_{\alpha = 1}^{2N_{p}} \hat{p}^{2}_{\alpha} + \omega^2_{\alpha}\hat{q}_{\alpha}^{2} \right) = \hat{T}_{p}(\dulq) + \hat{V}_{p}(\dulq).
\label{Hp}
\end{equation}
Here $\hat{q}_{\alpha} = \sum_{\alpha}\sqrt{\frac{\hbar}{2\omega_{\alpha}}}(\hat{a}^{+}_{\alpha} + \hat{a}_{\alpha})$ defines the photonic displacement coordinate for the $\alpha$th mode, with creation($a^+$) and annihilation($a$) operators \cite{Tokatly2013,Pellegrini2015}, and the commutation relation  $[\hat{q}_{\alpha},\hat{p}_{\alpha'}] = \imath\hbar\delta_{\alpha,\alpha'}$. The photonic displacement coordinate is directly proportional to the mode- projected electric displacement operator, ${\hat D}_{\alpha} = \epsilon_{0}\omega_{\alpha}\lambda_{\alpha}\hat{q}_{\alpha}$, while $\hat{p}_{\alpha}$ is proportional to the magnetic field \cite{Pellegrini2015,Flick2015}. The $\alpha$th mode has frequency $\omega_\alpha = k_\alpha c = \alpha\pi c/V$, with $k_\alpha$ the wavevector and $V$ the quantization volume. The electron-photon coupling strength is given by
\begin{align}
\label{eq:ab_initio_coupling}
\boldsymbol \lambda_\alpha= \sqrt{4 \pi} S_\alpha(\textbf{k}_\alpha\cdot \textbf{X})\textbf{e}_\alpha,
\end{align}
where $S_\alpha$ denotes the mode function, e.g. a sine-function for the case of a cubic cavity \cite{ruggenthaler2014quantum,Pellegrini2015}, $\textbf{k}_\alpha$ the wave vector,
and $\textbf{X}$ the total dipole of the system. In particular, we emphasize at this point that the mode functions introduce a dependence of the coupling constants on the quantization
volume of the electromagnetic field. By confining this volume, for example with an optical cavity, one can tune the interaction strength.
%FIXME: notation for the dipole clashes a bit with exact factorization notation.
 Finally, we note that the sum in Eq.~(\ref{Hp}) goes from 1 to $2N_{p}$, to take the two polarization possibilities of the electro-magnetic field into account. 
 The second term of Eq.~(\ref{fullH}) denotes the electronic Hamiltonian
\begin{align}
\hat{H}_{e}(\dulr) &= \sum_{i=1}^{N_{e}} \frac{\hat{\bp}^{2}_{i}}{2m_{e}} + \frac{e^{2}}{4\pi\epsilon_{0}}\sum_{i>j}^{N_{e}}\frac{1}{\vert \br_{i} - \br_{j}\vert}  \nonumber \\
%+ \frac{e^2}{2}\sum^{2N_p}_{\alpha = 1} \lambda_{\alpha}\left(\sum_{i} \br_{i}\right)^2
 &= \hat{T}_{e}(\dulr) + \hat{V}_{ee}(\dulr)\;,
 \label{He}
\end{align}
where $m_{e}$ defines the electronic mass, $\hat{\bp}_i$ the electronic momentum operator conjugate to $\hat\br_i$. 
The third term in Eq.~(\ref{fullH}) denotes the nuclear Hamiltonian
\begin{align}
\hat{H}_{n}(\dulR) &= \sum_{I=1}^{N_{n}} \frac{\hat{\bP_I}^{2}}{2M_{I}} + \frac{e^{2}}{4\pi\epsilon_{0}}\sum_{i>j}^{N_{n}}\frac{Z_IZ_J}{\vert \bR_{I} - \bR_{J}\vert}  \\
%+ \frac{e^2}{2}\sum^{2N_p}_{\alpha = 1} \lambda_{\alpha}\left(\sum_{I}Z_I \bR_{I}\right)^2 \nonumber
 &= \hat{T}_{n}(\dulR) + \hat{V}_{nn}(\dulR);,
 \label{Hn}
\end{align}
with analogous identifications to the electronic Hamiltonian and $eZ_I$ here being the nuclear charge.

The remaining terms in Eq.~(\ref{fullH}) denote the couplings between the subsystems. 
The electron-nuclear coupling appears as the usual Coulombic interaction:
\ben
\hat{H}_{en} = -\sum_{i=1}^{N_e}\sum_{J = 1}^{N_n}\frac{e^2 Z}{\vert \br_i - \bR_J\vert}
\label{Hen}
\een
The electron-photon coupling, in dipole approximation,
\begin{equation}
\hat{H}_{ep} = -\sum_{\alpha = 1}^{2N_{p}}\omega_{\alpha}\hat{q}_{\alpha}\vec{\lambda}_{\alpha}\cdot\sum_{i=1}^{N_{e}}e\br_{i} 
%= \hat{V}_{ep}(\dulr,\dulq),
\label{Hep}
\end{equation}
(where $e$ is the magnitude of the electronic charge) bilinearly couples the total electric dipole moment with the  electric field operator for each mode of the photonic field.  Similarly, the nuclear-photon coupling is
\begin{equation}
\hat{H}_{np} = \sum_{\alpha = 1}^{2N_{p}}\omega_{\alpha}\hat{q}_{\alpha}\vec{\lambda}_{\alpha}\cdot\sum_{I=1}^{N_{n}}eZ_I\bR_{I} 
%= \hat{V}_{ep}(\dulr,\dulq),
\label{Hnp}
\end{equation}
Finally, $H_{pen}$ represents the dipole self-energy of the matter in the radiation field:
\ben
\hat{H}_{pen} = \frac{1}{2}\sum_{\alpha = 1}^{2N_p}\vec\lambda_\alpha\cdot\left(\sum_I^{N_n}Z_I\bR_I - \sum_i^{N_e}\br_i  \right)^2
\een
This self-energy term is essential for a mathematically well defined light-matter interaction. Without this
term the Hamiltonian is not bound from below, and loses in addition translational invariance (in case of a vanishing
external potential) \cite{rokaj2017}.

The dynamics of such a coupled system is given by the solution of the time-dependent Schr\"odinger equation (TDSE)
\begin{equation}
\hat{H}\Psi(\dulr,\dulR,\dulq,t) = i\partial_{t}\Psi(\dulr,\dulR,\dulq,t),
\label{TDSE}
\end{equation}
where $\Psi(\dulr,\dulR,\dulq,t)$ is the full matter-photon wavefunction, that  contains the complete information of the coupled system. However it is difficult to obtain an intuitive understanding and interpretation of such a coupled system from the high-dimensional  $\Psi(\dulr,\dulR,\dulq,t)$, and moreover, we may not be interested in all the information as we might be interested in one of the subsystems. 
If one of these subsystems varies on a much slower time-scale than the others (in particular the nuclei), what is often done in coupled electron-nuclear systems is a Born- Oppenheimer adiabatic approximation where the faster \linebreak time-scale subsystem (in particular the electrons) are assumed to instantaneously adjust to the positions of the nuclei, and hence if they begin in an eigenstate, they remain in an eigenstate parameterized by the nuclear coordinate. The eigenenergy maps out  a Born-Oppenheimer potential energy surface (PES) which provides the potential for the nuclear dynamics. 
These potential-energy surfaces (PES) are clearly an approximation within the adiabatic ansatz, but in fact an {\it exact} PES can be defined quite generally without the need for any adiabatic approximation, which brings us to the main point of this paper. For the electron-nuclear problem, these arise from the exact factorization approach mentioned earlier in the introduction. In the next section we will extend the idea of the exact factorization for electron-nuclei systems to coupled photon-matter systems.

\begin{figure}[H]{
\includegraphics[width=0.6\linewidth]{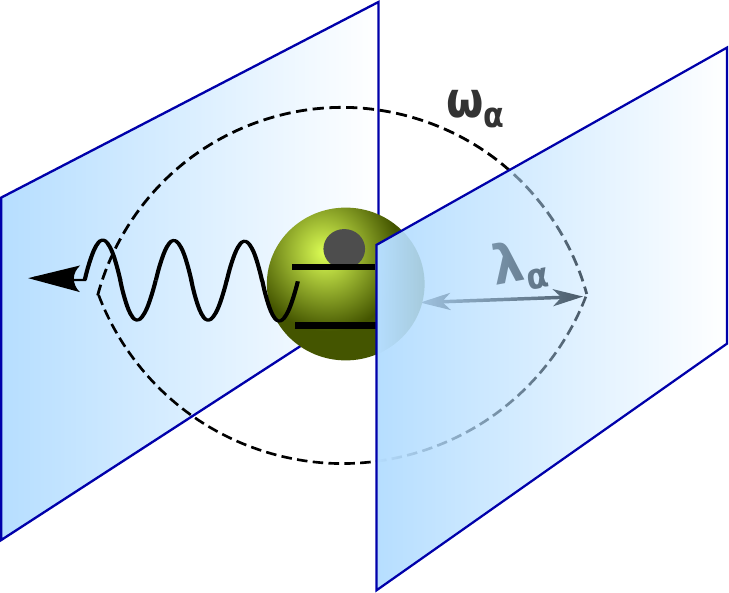}}
\caption{Cavity-setup: Particle (green) trapped in a cavity and coupled by coupling strength $\lambda_{\alpha}$ to the $\alpha$ photon mode with the photonic frequency $\omega_{\alpha}$, where $\alpha = \{1,2, ..., 2n_{p}\}$.}
\label{fig:AB1}
\end{figure}

Before moving to this, we note that Eq.~(\ref{fullH}) is the most general form of Hamiltonian that we will consider in the present work. In later sections, in particular in the explicit examples, we will simplify to just a two-level electronic system interacting with the photonic field in a cavity. In that case, many of the terms in Eq.~(\ref{fullH}) are zero, and we simplify the remaining terms even further to a model Hamiltonian 
\begin{equation}
\hat{H} = -\frac{\omega_{0}}{2}\hat{\sigma}_{z} + \sum_\alpha\left(-\frac{1}{2}\frac{\partial^2}{\partial q_\alpha^2} + \frac{1}{2}\omega^2_\alpha q_\alpha^2\right)
+\sum_{\alpha}\omega_{\alpha}\lambda_{\alpha}\hat{q}_{\alpha}(d_{eg}\hat{\sigma_{x}}),
%\sum_{\alpha}\omega_{\alpha}\hat{a}_{\alpha}^{+}\hat{a}_{\alpha}  
\label{Hmodel}
\end{equation}
Here $\sigma_i$ are the Pauli matrices. The first term is the  two-level system that replaces the electronic Hamiltonian (including the dipole self-energy, which simplifies to a constant energy shift for a two-level system), where the \linebreak energy-level difference is $\omega_0$, and $d_{eg}$, appearing in the third term, is the dipole moment of the transition. The second term describes the free photon field, as in Eq.~(\ref{Hp}), while Eq.~(\ref{Hep}) reduces to the third term with $\lambda_\alpha$ as the coupling strength evaluated at the position of the atom in the cavity. The TDSE also simplifies, to
\ben
i\hbar \frac{\partial}{\partial t} \orr{\Psi}(\dulq, t)= \begin{pmatrix}-\frac{\omega_0}{2} + \hat{H}_p(\dulq) & \sum_\alpha\omega_{\alpha}\lambda_{\alpha}\hat{q}_{\alpha}d_{eg} \\ \sum_\alpha\omega_{\alpha}\lambda_{\alpha}\hat{q}_{\alpha}d_{eg} & \frac{\omega_0}{2}+ \hat{H}_p(\dulq) \end{pmatrix}\orr{\Psi}(\dulq,t)
\label{TDSE_tls}
\een
where we use the notation $\orr{\Psi}(\dulq,t)$ being a 2-vector defined at every $\dulq$ and $t$. A cartoon of the problem is given in Fig.~\ref{fig:AB1}. 

\subsection{Exact Factorization Approach}
\label{sec:EFA}
The exact factorization (EF) may be viewed as a reformulation of the quantum mechanics of interacting coupled systems where the wavefunction is factored into a marginal amplitude and a conditional amplitude~\cite{H75,GG14,AMG10,AMG12}. With non-relativistic electron-nuclear systems in mind, the equations for these amplitudes were derived for Hamiltonians of the form 
\ben
\hat{H} = \hat{T}_e + \hat{T}_n + \hat{V}
\label{H_EFA}
\een
where $\hat{V}$ is a scalar potential that includes coupling between the electrons and nuclei (usually  Coulombic) and any externally applied fields. Here $\hat{T}_{e,n}$ are kinetic energy operators of the electronic and nuclear equations, just as in Eq.~\ref{He} and~\ref{Hn}, that have the form of $-\sum_{i(I)}\nabla^2_{i(I)}/2m_{i(I)}$ (that is, no vector potential). The EF then proves that the exact full molecular wavefunction can be factored as 
\ben
\Psi(\dulr,\dulR,t)  = \chi(\dulR,t)\Phi_\dulR(\dulr,t)\;.
\label{EFAorig}
\een
The equation for the nuclear amplitude $\chi$ has a TDSE form~\cite{AMG10,AMG12,AMG13,ACEJ13}, equipped with a time-dependent scalar potential $\epsilon(\dulR, t)$ and a time-dependent vector potential  $\bA_I(\dulR,t)$ that include entirely the effects of coupling to the electronic system as well as external fields. The equation for the conditional electronic amplitude $\Phi_\dulR$ has a more complicated form, involving a coupling operator $\hat{U}_{en}$, that acts on the parametric dependence of $\Phi_\dulR$. The factorization is unique, provided $\Phi_\dulR$ satisfies the "partial normalization condition" (PNC), $\int d\dulr \vert \Phi_\dulR(\dulr,t)\vert^2 = 1$, up to a gauge-like transformation; under such a transformation, $\epsilon$ and $\bA$ transform as scalar and vector potentials do in electrodynamics.   The nuclear $N_n$-body probability density and current-density can be obtained in the usual way from the nuclear amplitude $\chi(\dulR, t)$, so in this sense, can be identified as the nuclear wavefunction of the system.

The form of the EF Eq.(\ref{EFAorig}) is  similar to the Born- Oppenheimer (BO) approximation, however with the important difference that Eq.(\ref{EFAorig}) is an {\it exact} representation of the wavefunction, not an approximation, and further that it is valid for time-dependent systems, with time-dependent external fields, as well. The BO approximation assumes that the electronic system remains always in the instantaneous ground (or eigen)-state associated with the nuclear configuration $\dulR$, and therefore misses all the physics associated with non-adiabatic effects, including wavepacket branching and decoherence. These effects are contained exactly in the coupling terms in the EF equations: the scalar and vector potentials and the coupling operator of the electronic equation. It is important to note that there is no assumption of different \linebreak timescales in the EF approach, in contrast to the BO approximation. 

As the scalar potential plays a role analogous to the BO PES, but now for the exact system, it is denoted the time-dependent potential energy surface (TDPES), while the vector potential (TDVP) is an exact time-dependent Berry connection. The gauge-freedom is a crucial part of the EF approach: in particular, whether a gauge exists in which the vector potential can be transformed into part of the TDPES has been explored in some works~\cite{MAKG14,RTG16}, especially since the common understanding is that Berry phases appear only out of an adiabatic separation of time-scales, while the EF is exact and does not assume any such separation. 
In fact, equally valid is the reverse factorization~\cite{SAMYG14},  $\Psi(\dulr,\dulR,t)  = \chi(\dulr,t)\Phi_\dulr(\dulR,t)$, which is particularly useful when one is interested in the electronic system, since in this factorization, the electronic system follows a TDSE in which the potentials can be analysed and interpreted.

\subsection{Exact Factorization Approach for QED}
\label{sec:EFA-QED}
Here, we extend the exact factorization to systems of coupled photons, electrons, and nuclei. Since all the kinetic operators in the Hamiltonian within the dipole approximation, Eq.~(\ref{fullH}), are of similar form to those that were considered in the original EFA, Eq.~(\ref{H_EFA}), the mathematical structure of the equations and coupling terms will be similar when we make a factorization into two parts.

There are three possibilities for such a factorization, and we expect each to be useful in different contexts. 
One possibility, which is perhaps the most natural extension of the factorization of Ref.~\cite{AMG10,AMG12}, is to take the nuclear system as the marginal one, 
\ben
\Psi(\dulq,\dulr,\dulR; t) = \chi(\dulR; t)\Phi_{\dulR}(\dulq,\dulr;t)\;.
\label{EFmargR}
\een
with the PNC
\ben
\langle \Phi_{\dulR}(t) \vert \Phi_\dulR(t)\rangle_{\dulq,\dulr} \equiv \int d\dulq d\dulr \vert\Phi_\dulR(\dulq,\dulr;t)\vert^2 = 1 \een
for every nuclear configuration $\dulR$ at each time $t$. 
This would yield a TDSE for the nuclear system, much like in the original EFA, but now the TDPES and TDVP includes not only the effects on the nuclei of coupling to the electrons, but also to the photons.  This would be a particularly useful factorization for studying light-induced non-adiabatic chemical dynamics phenomena, when the quantum nature of light is expected to play a role. In fact, an approximation  based on the normal  Born- Oppenheimer approximation for the electron-ion dynamics has been used  to study the cavity-induced changes in the potential energy surfaces in the strong coupling regime ~\cite{GGF15}. This would be a particularly useful factorization for studying light-induced non-adiabatic phenomena, when the quantum nature of light is expected to play a role. 

A second possibility is the natural extension of the reverse factorization~\cite{SAMYG14}, where the electronic system is the marginal amplitude
\ben
\Psi(\dulq,\dulr,\dulR; t) = \chi(\dulr; t)\Phi_{\dulr}(\dulq,\dulR;t)\;,
\label{EFmargr}
\een
with the PNC $\langle \Phi_{\dulr}(t)\vert \Phi_\dulr(t)\rangle_{\dulq,\dulR} \equiv \int d\dulq d\dulR \vert\Phi_\dulr(\dulq,\dulR;t)\vert^2 = 1$, for all $t$ and every electronic configuration $\dulr$,
which would yield a TDSE for electrons, with the e-TDPES and e-TDVP now incorporating the full effects on the electrons of coupling to the nuclei as well as the photons. This could be particularly useful for studying, for example,  the impact of vacuum field on electrical conductivity in a molecule or semiconductor. 

This leaves the third possibility, where the photonic system is chosen as the marginal: 
\ben
\Psi(\dulq,\dulr,\dulR; t) = \chi(\dulq; t)\Phi_{\dulq}(\dulr,\dulR;t)\;,
\label{EFmargq}
\een
with the PNC 
\ben
\langle \Phi_{\dulq} (t)\vert \Phi_\dulq(t)\rangle_{\dulr,\dulR} \equiv \int d\dulr d\dulR \vert\Phi_\dulq(\dulr,\dulR;t)\vert^2 = 1\,,
\label{PNCmargq}
\een
for each field-coordinate $\dulq$ and all times $t$. 
This is the factorization we will focus on in the present paper: it gives a TDSE for the photonic system, within which the scalar potential, which we call the q-TDPES, and vector potential, the q-TDVP, contain the feedback of the matter-system on the radiation field. In free space, the potential acting on the photons is quadratic  as is evident from Eq.~(\ref{Hp}), however, in the presence of matter, the potential determining the photonic state deviates from its harmonic form due to interactions with matter. The cavity-BO approach introduced in Ref.~\cite{Flick2017a} has demonstrated these deviations within the BO approximation. The EF approach now renders this concept exact, beyond any adiabatic assumptions.

The equations for each of these three factorizations follows from a straightforward generalization of the original EF equations, as the non-multiplicative operators (the kinetic operators) have the same form; hence the derivation proceeds quite analogously to that given in Ref.~\cite{AMG10,AMG12,AMG13}. In particular, for the factorization Eq.~(\ref{EFmargq}), we obtain
\begin{align}
&\left(\hat{H}_{m}(\dulr,\dulR,\dulq;t) - \epsilon(\dulq;t)\right)\Phi_{\dulq}(\dulr,\dulR;t) = i\partial_{t}\Phi_{\dulq}(\dulr,\dulR;t),\label{EFeqnsmargq}\\
&\left(\sum_{\alpha}^{2n_{p}}\frac{1}{2}\Big(i\frac{\partial}{\partial_{q_\alpha}} + A_{\alpha}(\dulq;t)\Big)^2 + \epsilon(\dulq;t)\right)\chi(\dulq;t) = i\partial_{t}\chi(\dulq;t),
\end{align}
where the matter Hamiltonian $\hat{H}_m$ is given by 
\begin{equation}
\hat{H}_{m}(\dulr,\dulR,\dulq;t) =\hat{H}_{\rm qBO} +  \hat{U}_{ep}.
\label{Hm}
\end{equation}
with 
\ben
\hat{H}_{\rm qBO} = \hat{H}_e+\hat{H}_n+\hat{H}_{en} +\hat{H}_{pen} +\hat{H}_{ep} +\hat{H}_{np}+ \frac{1}{2}\sum_{\alpha = 1}^{2N_p}\omega_\alpha\hat{q}^2_\alpha
\label{HqBOmargq}
\een
defined in an analogous way to the BO Hamiltonian, but now for the photonic system. 
The electron-photon coupling potential $\hat{U}_{ep}$ is given by
\begin{align}
&\hat{U}_{ep}[\Phi_\dulq,\chi] = \sum_{\alpha}^{2n_{p}} \Bigg[ \frac{(-i\partial_{q_\alpha} - A_{\alpha}(\dulq;t))^2}{2} + \label{G13} \\
&\left(\frac{-i\partial_{q_\alpha}\chi(\dulq;t)}{\chi(\dulq;t)} + A_{\alpha}(\dulq;t)\right)\left( -i\partial_{q_\alpha} - A_{\alpha}(\dulq;t)\right)\Bigg] 
\nonumber,
\label{Uep}
\end{align}
the q-TDPES by
\begin{equation}
\epsilon(\dulq;t) = \Bra{\Phi_{\dulq}(t)}\hat{H}_m(\dulr,\dulR,\dulq;t) - i\partial_{t}\Ket{\Phi_{\dulq}(t)}_{\dulr,\dulR},\label{qtdpes}
\end{equation}
and the q-TDVP by 
\begin{equation}
A_{\alpha}(\dulq;t) = \braket{\Phi_{\dulq}(t)|-i\partial_{q_\alpha}\Phi_{\dulq(t)}}_{\dulr,\dulR}.\label{qTDVP}
\end{equation}
with the notation $\langle ...\vert..\rangle_{\dulr,\dulR}$ meaning an integration over electronic and nuclear coordinates only in the expectation values (as in Eq.~(\ref{PNCmargq})).
The factorization~(\ref{EFmargq}) is unique up to a gauge-like transformation, provided the PNC, Eq.~(\ref{PNCmargq}) is satisfied. 
The gauge-like transformation has the structure of the usual one in electromagnetism, except here the scalar and vector potentials arise due to coupling, rather than due to external fields, and they are potentials on the photonic system, not on the matter system. The equations are form-invariant under the following transformation: 
\bea
\nonumber
\Phi_\dulq(\dulr, \dulR, t) &\to& \Phi_\dulq(\dulr, \dulR, t) \exp(i \theta(\dulq, t))\\
\nonumber
\chi(\dulq,t) &\to& \chi(\dulq,t)\exp(-i\theta(\dulq,t))\\
\nonumber
A_\alpha(\dulq; t) &\to&A_\alpha(\dulq; t) + \partial_\alpha\theta(\dulq, t)\\
\epsilon(\dulq;t) &\to& \epsilon(\dulq; t) + \partial_t\theta(\dulq, t)
\eea

Further, one can show that the displacement-field density represented by $\chi$ reproduces that of the full wavefunction, i.e.  
\ben
\vert \chi(\dulq ;t)\vert^2 = \int d\dulr d\dulR \vert\Psi(\dulq,\dulr,\dulR;t)\vert^2\,,
\een
 and that the phase of $\chi$ together with the q-TDVP provide the displacement-field probability current in the natural way: 
 \ben
 \mbox{Im} \langle\Psi\vert\partial_\alpha\Psi\rangle =\vert\chi(\dulq;t)\vert^2\bA_\alpha(\dulq;t) + \partial_\alpha S(\dulq;t) \,,
 \een
where $\chi(\dulq;t) = \vert\chi(\dulq;t)\vert \exp(i S(\dulq, t))$. 
This means, that observables associated with multiplication by $\dulq$ can be obtained directly from $\chi(\dulq, t)$, for example, the electric field 
\ben
{\bf{E}}(\br;t)  = \sum_\alpha \omega_\alpha  {\boldsymbol{\lambda}}_\alpha(\br,t)\int d\dulq \,q_\alpha \vert\chi(\dulq; t)\vert^2\,,
\een
while the magnetic field is 
\ben
{\bf{B}}(\br;t)\! =\! \sum_\alpha \frac{c}{\omega_\alpha} \nabla \times \boldsymbol{\lambda}_\alpha(\br,t)\!\!\int\!\! d\dulq \vert\chi(\dulq;t)\vert^2 A_\alpha(\dulq;t) + \partial_\alpha S(\dulq;t).
\een

\subsection{Exact Factorization for Simplified Model Hamiltonian: Two-Level System in Radiation Field}
\label{sec:EFA_model}
For our exploration of the QED factorization in this paper, we will turn to the simplified model Hamiltonian of Eq.~(\ref{Hmodel}), where the matter system's Hamiltonian is a 2$\times$2 matrix. 
First, it is useful to write Eq.(~\ref{Hmodel}) as
\bea
\hat{H} &=& \sum_\alpha\frac{1}{2}\partial_{q_\alpha}^2\mathds{1}_2 + \hat{H}_{\rm qBO}\,, {\rm where}\\
\hat{H}_{\rm qBO}\! &=&\! -\frac{\omega_{0}}{2}\hat{\sigma}_{z} \! +\!\! \sum_\alpha \!\frac{1}{2}\omega^2_\alpha q_\alpha^2\mathds{1}_2
\! +\!\! \sum_{\alpha}\!\omega_{\alpha}\lambda_{\alpha}\hat{q}_{\alpha}(d_{eg}\hat{\sigma_{x}}).
\label{HqBO}
\eea
Here $\hat{H}_{\rm qBO}$ is analogous to the BO Hamiltonian in the usual electron-nuclear case. We can define qBO-states as normalized eigenstates:
\ben
\hat{H}_{\rm qBO}\overrightarrow{\Phi}_\dulq^{(1,2)} = \epsilon^{(1,2)}_{\rm qBO}(\dulq)\orr{\Phi}_\dulq^{(1,2)}
\label{HqBOeqn}
\een
with $\orr{\Phi}^{i,\dagger}_\dulq\cdot\orr{\Phi}_\dulq^{j} = \delta_{ij}$.
and these can be used as a basis to expand the fully coupled wavefunction, i.e. 
\ben
\orr{\Psi}(\dulq, t) = \chi_1(\dulq, t) \orr{\Phi}_\dulq^{(1)} + \chi_2(\dulq, t) \orr{\Phi}_\dulq^{(2)} 
\label{BHexp}
\een
which would be analogous to the Born-Huang expansion but now for the cavity-matter system.

Now in the EF approach, the fully coupled wavefunction is  instead factorized as a single product:
\ben
\orr{\Psi}(\dulq, t) = \chi(\dulq, t) \orr{\Phi}_\dulq(t)
\een
where the PNC becomes 
\ben
\orr{\Phi}^\dagger_\dulq(t)\cdot\orr{\Phi}_\dulq(t) = 1\;,
\label{PNCmodel}
\een
and holds for every $\dulq$ and each time $t$. 

We note that there are two useful bases for this problem. One is obtained from diagonalizing the field-free two-level system, i.e. that defined by eigenvectors of the Pauli-$\sigma_z$ matrix. The other basis is the qBO basis, defined by the eigenvectors of $\hat{H}_{\rm qBO}$, as in Eq.(~\ref{HqBOeqn}). 

The EF equations follow directly from Eqs~(\ref{EFeqnsmargq}--\ref{qTDVP}) but with the much simplified $\hat{H}_{\rm qBO}$ above, and all $\langle ...\rangle$ meaning simply a 2$\times$2 vector-multiply. 

\subsection{Photonic Time-Dependent Potential Energy Surface}
Unlike the original electron-nuclear factorization, the q-TDVP can always be gauged away due to the one- dimensional nature of each photon-displacement mode. \linebreak This means that one can always transform to a gauge in which the q-TDPES contains the {\it entire} effect of the coupling of the matter system on the radiation field, i.e. it is the only potential that is driving the photonic dynamics. For the matter system, both the q-TDPES and the photon-matter coupling operator incorporate the effect of the photonic system on the matter.
In the original electron-nuclear factorization, the TDPES proved to be a powerful tool to analyze and interpret the dynamics of the system in cases ranging from dynamics of molecules in strong fields~\cite{AMG10,AMG12,SAMYG14,SAMG15,KAM15,FHGS17,SG17}, non-adiabatic proton-coupled electron- transfer~\cite{AASG13,AASMMG15} to nuclear-velocity perturbation theory ~\cite{SASGV17,SASGV15} and dynamics through a conical intersection~\cite{HAGE17,CA17}. It provides an exact generalization of the adiabatic BO-PES. 

In the present work, we will study the q-TDPES $\epsilon(\dulq,t)$ of Eq.(\ref{qtdpes}) for the case of the radiation field coupled to a two level-system, using the model Hamiltonian~(\ref{Hmodel}). Given a solution $\orr{\Psi}(\dulq,t)$ for the coupled system, found from Eq.~(\ref{TDSE_tls}), we will extract the  exact q-TDPES by inversion.

To do this, we first ensure that we work in the gauge where $A_\alpha = 0$. Similarly to previous work~\cite{AMG12,AASMMG15}, this gauge can be fixed by choosing the phase $S(\dulq, t)$ of the photonic wavefunction,  $\chi(\dulq, t) = \vert \chi(\dulq,t)\vert \exp(i S(\dulq, t))$, to satisfy
\begin{equation}
\partial_{q_\alpha} S(\dulq,t) =\frac{\text{Im}[\braket{\orr{\Psi}(t)|\partial_{q_\alpha}\orr{\Psi}(t)}]}{|\chi(\dulq,t)|^2}. 
\label{chi_phase}
\end{equation}
Then, from the given solution $\orr{\Psi}(\dulq, t)$, we compute 
\ben
\orr{\Phi}_{\dulq}(\dulr) = \frac{\orr{\Psi}(\dulq,t)}{\vert\chi(\dulq,t)\vert e^{iS(\vec{q,t})}}\, {\rm with } \vert\chi\vert = \sqrt{\orr{\Psi}^\dagger(\dulq, t)\cdot\orr{\Psi}(\dulq, t)}
\label{phiq}
\een
and insert into the q-TDPES
\begin{align}
\nonumber
\epsilon(\dulq,t) &= \orr{\Phi}^\dagger_{\dulq}(t)\cdot\hat{H}_{\rm qBO}\cdot\orr{\Phi}_\dulq(t)\\
\nonumber
 &+ \sum_\alpha \vert\partial_\alpha\orr{\Phi}_\dulq(t)\vert^2 + \orr{\Phi}_\dulq^\dagger(t)\cdot (-i\partial_t \orr{\Phi}_\dulq(t))\\
%\Bra{\orr{\Phi}_{\dulq}(t)}\hat{H}_{BO}(\dulq,\dulr,t)\Ket{\orr{\Phi}_{\dulq}(t)}  \label{eps_model}\\
%&+\left\langle\orr{\Phi}_{\dulq}(t)\vert\sum_{\alpha}^{2n_{p}}\frac{\partial_{q_\alpha}^2}{2}\orr{\Phi}_{\dulq}(t)\right\rangle+ \langle\orr{\Phi}_{\dulq}(t)\vert -i\partial_{t}\orr{\Phi}_{\dulq}(t)\rangle\nonumber\\
&= \epsilon_{\rm wBO}(\dulq,t) + \epsilon_{\rm kin}(\dulq,t)+ \epsilon_{\rm GD}(\dulq,t) \;. 
\label{eps_model}
\end{align} 
where we have noted that in this gauge the electron-photon coupling operator reduces to
${U}_{ep} = \sum_{\alpha}^{2n_{p}}\partial^2_{\alpha}/2$.
We have identified here the first term in Eq.~(\ref{eps_model}) as $\epsilon_{\rm wBO}$, a "weighted qBO" surface, weighted by the probabilities of being in the qBO eigenstates: using the expansion Eq.~(\ref{BHexp}), $\epsilon_{\rm wBO} = \left(\vert\chi_1(\dulq,t)\vert^2\epsilon_{\rm qBO}^{(1)}(\dulq, t) + \vert\chi_2(\dulq,t)\vert^2\epsilon_{\rm qBO}^{(2)}(\dulq, t)\right)/\vert\chi(\dulq t)\vert^2$. The second term arises from kinetic effects from the parametric dependence of the conditional matter wavefunction, hence we denote it as $\epsilon_{\rm kin}$.  Both those terms are invariant under different gauge choices, while the last term is gauge-dependent, hence its name $\epsilon_{\rm GD}$.

\section{Results and Discussion}
\label{sec:results}
We will consider two extremes within the simplified \linebreak model Hamiltonian Eq.~(\ref{Hmodel}). The first is the Wigner- \linebreak Weisskopf limit where the two-level system is coupled to an infinite number of cavity modes. The second is the two-level system coupled to a single resonant mode. As initial condition, we consider the photon modes in the vacuum state, and the two level system in the excited state. In the latter case, we will compare the effect of starting in a purely factorized matter-photonic state with that of starting in a qBO state.

We notice that the dipole matrix element and coupling parameter appear only together as a product in this model, $d_{eg}\lambda$. Physically, these are fixed by the problem at hand, specifically the volume of the cavity and the dipole coupling between the two levels in the atom, apart from fundamental constants. But here, in this model we choose them arbitrarily, and compare dynamics for different $d_{eg}\lambda$ that range from relatively weak coupling to strong coupling.

\subsection{Wigner-Weisskopf Limit}
\label{sec:WW}
We first consider the Wigner-Weisskopf limit, in which our two-level system is coupled to an infinite number of modes.
In this limit, the accepted well-known approximate solution for the coupled system is known analytically, which makes the q-TDPES particularly straightforward to find.

The solution for $\orr{\Psi}$ of the coupled problem can be found in the standard literature~\cite{ScullyZubairy}. The initial state is taken to be a purely factorized state of the electron in the excited state and all photon modes in their ground states, i.e.
\begin{equation}
\orr{\Psi}(\dulq,0) =\chi_{0}(\dulq) \begin{pmatrix}1\\0\end{pmatrix},
\label{G27}
\end{equation}
where 
\ben
\chi_{0}(\dulq) = \prod_{\alpha} \left(\frac{\omega_{\alpha}}{\pi\hbar}\right)^{\frac{1}{4}}e^{-\omega_{\alpha}q^{2}_{\alpha}/2\hbar}\;,
\een
which follows from the harmonic nature of the free photon field. 
The coupling in the off-diagonal elements of Eq.~(\ref{Hmodel}) then cause $\Psi$ to evolve in time, as 
\begin{equation}
\orr{\Psi}(\dulq, t) = a(t)\chi_{0}(\dulq)\begin{pmatrix}1\\0\end{pmatrix} + \sum_{\alpha}b_{\alpha}(t)\chi_\alpha(\dulq)\begin{pmatrix}0\\1\end{pmatrix} 
\label{G28}
\end{equation}
 under the reasonable assumption that the coefficients of the two-photon and higher states are negligible. Here the one-photon states of the photonic system are
\ben
 \chi_{\alpha}(\dulq) = \sqrt{\frac{2\omega_{\alpha}}{\hbar}}q_{\alpha}\prod_{\beta} \left(\frac{\omega_{\beta}}{\pi\hbar}\right)^{\frac{1}{4}}e^{-\omega_{\beta}q^{2}_{\beta}/2\hbar}\;.
 \een
 The coefficients $a(t)$ and $b_\alpha(t)$ can be found by substituting Eq.~(\ref{G28}) into the TDSE Eq.~(\ref{TDSE_tls}). After making the Wigner-Weisskopf approximations (taking the continuum limit so $V \to \infty$, taking $a(t)$ to change with a rate much slower than the resonant frequency $\omega_0$ and performing a Markov rotating-wave approximation, and neglecting a divergent Lamb shift), we arrive at
 \begin{align}
a(t) &= e^{-\frac{i\omega_{0}t}{\hbar}}e^{-\frac{\Gamma t}{2}},\\
b_{\alpha}(t) &= e^{i\omega_{\alpha}}\frac{ig_{\alpha}(e^{i(\omega_{\alpha} - \omega_{0})t - \Gamma t/2} - 1)}{i(\omega_{\alpha} - \omega_{0}) - \Gamma /2}.
\end{align}
where $g_{\alpha} = \sqrt{\frac{\pi\omega_{\alpha}}{2\hbar}}\lambda_{\alpha}d_{eg}$ and the decay (spontaneous emission rate), $\Gamma = (d_{eg}\lambda)^2\omega_0^2 \frac{V}{\hbar c^3}$.
The Wigner-Weisskopf 
solution is accurate for weak coupling, so that in this limit the solution also generates accurate q-TDPES.

With this Wigner-Weisskopf solution, we can then find the corresponding "exact" q-TDPES, Eq.~(\ref{eps_model}), using \linebreak Eq.~(\ref{phiq}) and~(\ref{chi_phase}). However, this yields an infinite dimensional surface, since $\dulq = (q_1..q_\alpha...q_ \infty)$, which is challenging to visualize. Instead, we plot some one-dimensional cross-sections of the q-TDPES, along the $i$th mode, setting $q_{\alpha\neq i} = 0$. In the following, we use $\bar{q}$ to denote all modes not equal to $q_i$. We will abbreviate quantities such as $\epsilon(q_i,\bar{q} =0,t)$ by $\epsilon(q_i,t)$, understood to be looking at the cross-section where the displacement-coordinate of all other modes is zero. We will choose two different \linebreak modes to look along: one resonant with $\omega_0$, and the other slightly off-resonant. 
With this choice of cross-sections through the origin of all modes but one, it can be shown that the phase of the nuclear wavefunction that satisfies the zero-q-TDVP condition, Eq.~({\ref{qTDVP}}),  $S(q_i,t) \equiv 0$. This leads to some simplification in the components of $\epsilon(q_i,t)$.

\begin{figure}[H]
 \centering
\includegraphics*[width=1\columnwidth]{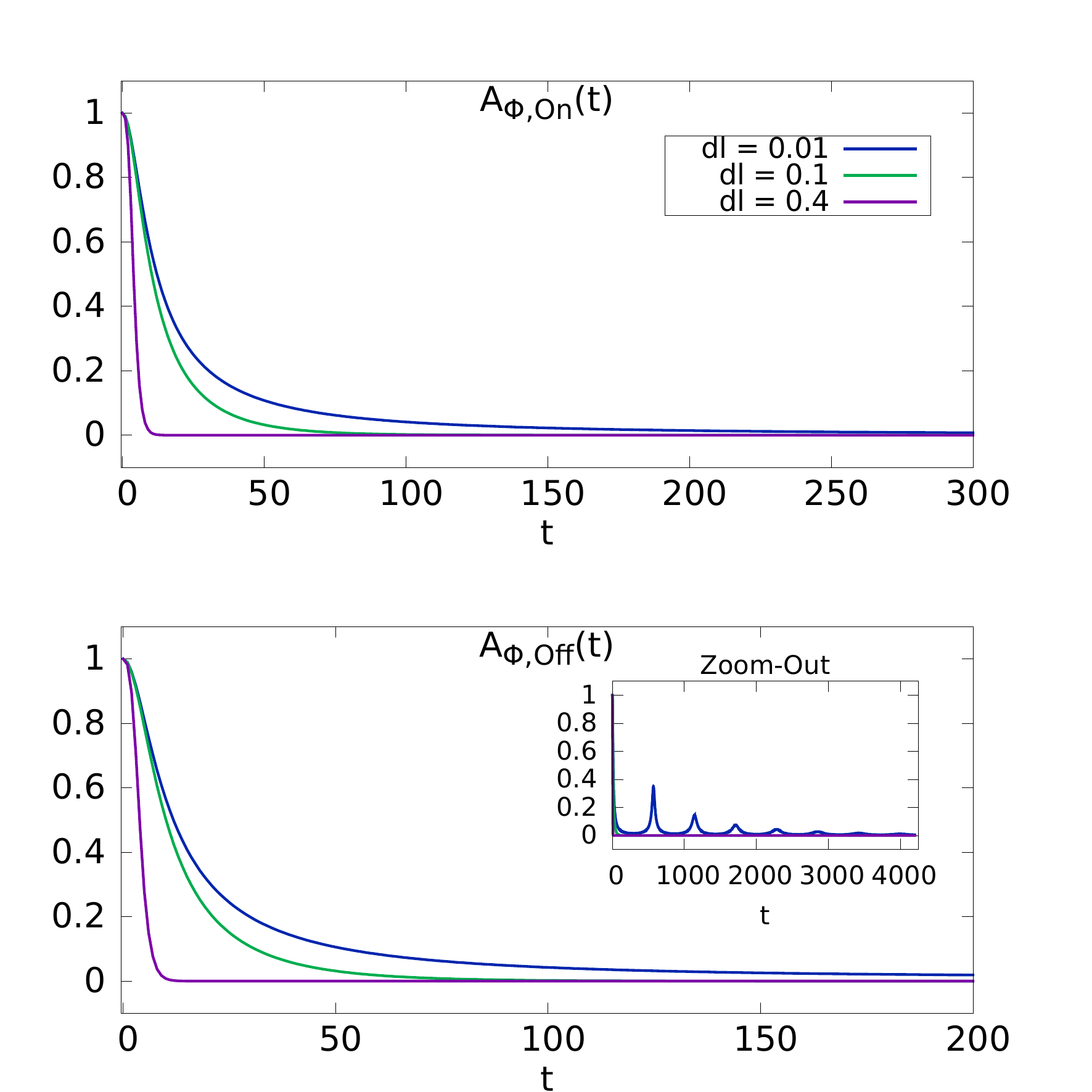}
 \caption{The autocorrelation function $A_{\Phi}(t)$ %$\int dq_i \orr{\Phi}^\dagger(q_i,0)\cdot\orr{\Phi}(q_i,t)$}
where the different colors describe the different coupling strengths of the system.  The upper panel shows
the decay of this function when we choose to look on-resonance $\omega_i = \omega_0 = 0.4$.
 The lower panel illustrates the decay when looking along a slightly-off resonant mode $\omega_{i} =\omega_{0} + 0.01$. The zoom-out shows the same off-resonance decay for a longer time.}
  \label{F1}
  \end{figure}

In Fig.\ref{F1} we plot the autocorrelation function 
\begin{equation}
A_{\Phi}(t) = \Bigl\lvert\int dq_i  (\orr{\Phi}^\dagger(q_i; t=0)\cdot\orr{\Phi}(q_i; t))\Bigr\rvert^2.
\end{equation}
as this gives an indication of what to expect for time-scales for
 the behavior of the q-TDPES $\epsilon(\vec{q_i},t)$ for different coupling strengths $d_{eg}\lambda= \{0.01,0.1,0.4\}$. In the upper panel, we have chosen to plot the q-TDPES along the mode of the radiation field that is resonant with the two-level system. In this case, the decay of the autocorrelation  depends primarily on $(d_{eg}\lambda)^2$, through $\Gamma$, i.e. $A_{\Phi}(t) \propto e^{-\Gamma t}$, although there are some small polynomial corrections.
 
\begin{figure}
 \centering
\includegraphics*[width=1\columnwidth]{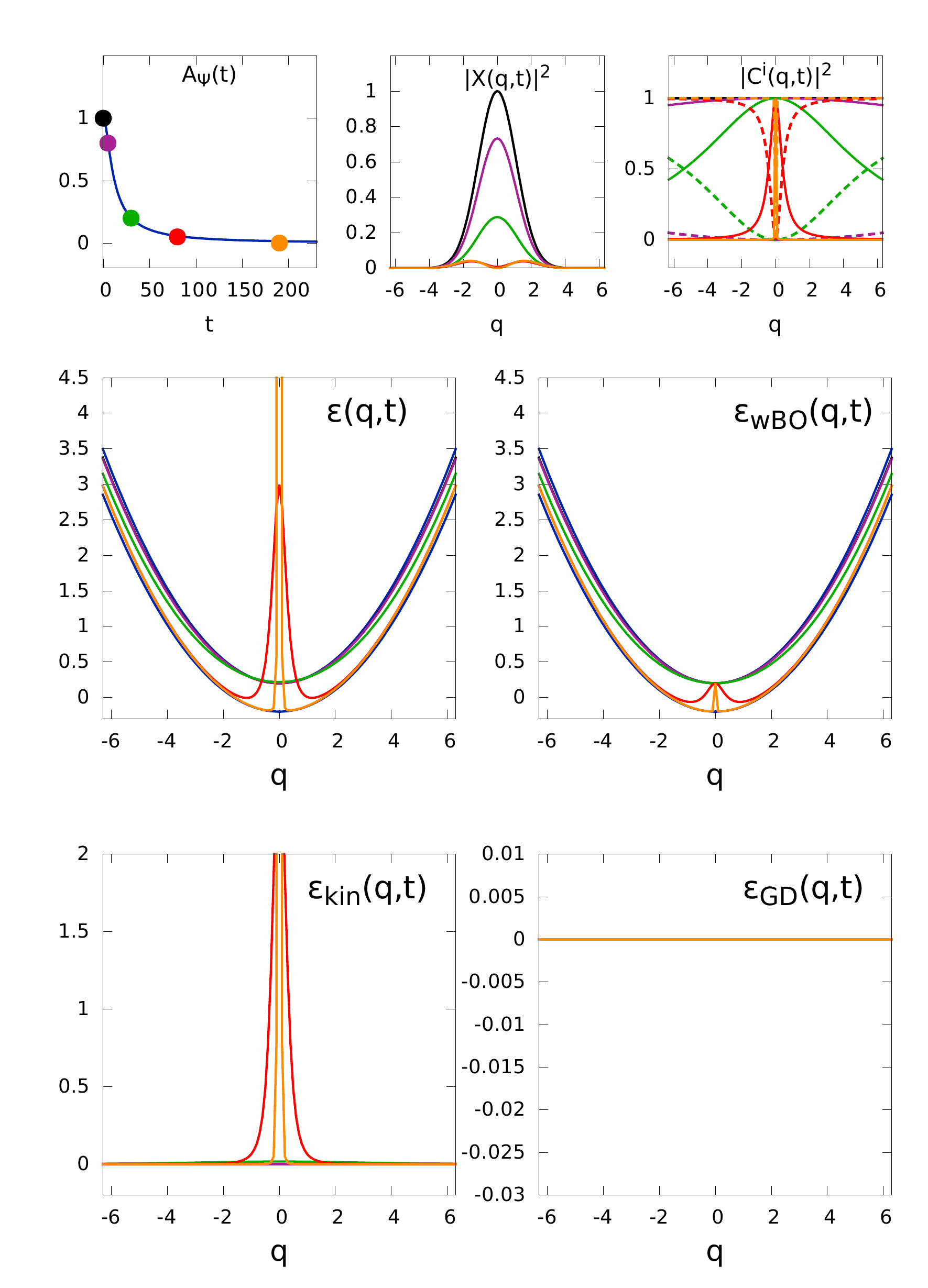}
 \caption{Wigner-Weisskopf model, looking on-resonance  $\omega_{0} = \omega_{k}$. The top left panel shows the decay $A_{\Phi}(t)$, where the different colored dots depict the times of the different time snapshots of the dynamics shown within this plot. The middle and right panels along the top show the photonic distribution $\vert\chi(\vec{q},t)\vert^2$ and each coefficient of the conditional electronic distribution $|C^{(1)}(q_i;t)|^2$ (dashed), $|C^{(2)}(q_i;t)|^2 $ (solid) at the corresponding time snapshots. The middle and lower panels show the different components of the $\epsilon(q_i;t)$ as well as the full scalar potential at the given time snapshots. In the middle panels, the qBO surfaces are shown in blue for reference. }
  \label{fig:WW-on}
\end{figure}

\begin{figure}
 \centering
\includegraphics*[width=1\columnwidth]{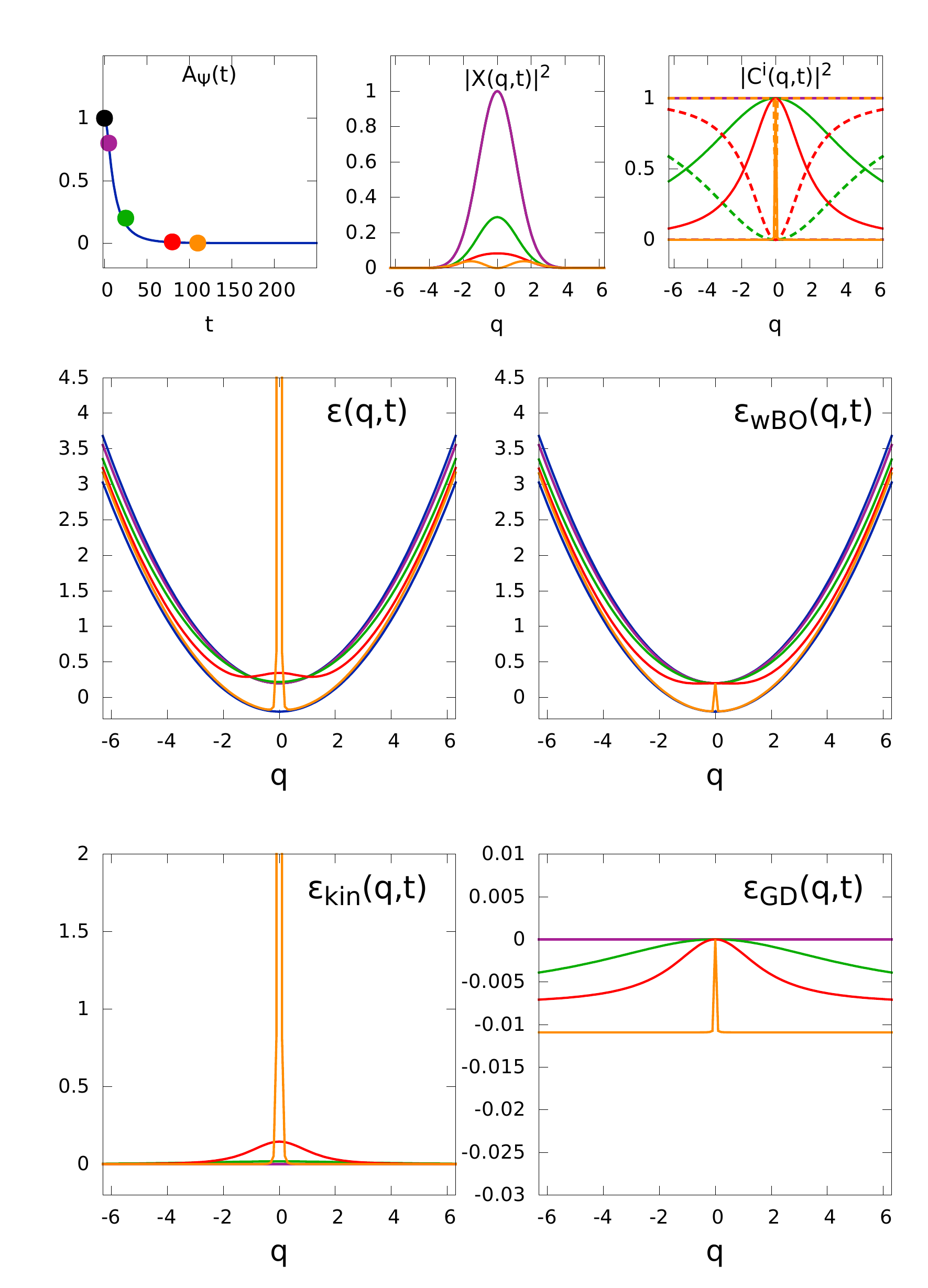}
 \caption{As for Figure~\ref{fig:WW-on} but looking along the slightly-off resonant mode in the Wigner-Weisskopf model.}
  \label{fig:WW-off}
\end{figure}

 In the slightly off-resonant case, we have chosen $\omega_i = 0.411$  while $\omega_0 = 0.4$. In fact, we observe partial revivals in in the autocorrelation function for very long times in the case of the weakest coupling shown ($d_eg\lambda = 0.01$), as shown in the inset, with the amplitude decreasing with each revival.  However the initial decay follows a similar $d_{eg}\lambda$-scaling pattern to that of the on-resonant section through the autocorrelation function. 
In either case, the dynamics of the  decay is essentially the same for all coupling strengths, provided the time is scaled appropriately, and their q-TDPES's also map on to each other at the corresponding times. In the following we show the graphs for $d_{eg}\lambda = 0.01$ for the cross-section taken along the on-resonant mode in Fig.~\ref{fig:WW-on}, and that along the off-resonant mode in Fig.~\ref{fig:WW-off}.

In Fig. \ref{fig:WW-on} and Fig.\ref{fig:WW-off} we show the different components of the q-TDPES $\epsilon(q_i,t)$ for different time snapshots which are illustrated by the colored dots in the decay-plot. The displacement-field density, $\vert\chi(q_i,t)\vert^2 = \vert\chi(q_i,\bar{q}=0,t)\vert^2$ at these time-snapshots is shown in the top middle panel, and we observe the gradual evolution from the vacuum state towards the state with one photon during the decay. This is also seen in the conditional probability amplitudes shown in the top right panel, which we obtain from 
\ben
\vert C^{1(2)}(q_i,t)\vert^2 = \orr{\Phi}_{q_i,\bar{q} =0}^{(1(2))}\cdot\orr{\Phi}_{q_i,\bar{q} =0}(t)
\een
These are the coefficients of expansion of $\Phi_{q_i}(t)$ in the BO basis and are equal to the coefficients in Eq.~(\ref{BHexp}) via $C^j(q_i,t) = \chi_j(q_i,t)/\chi(q_i,t)$. 
$C^{(1)}(q_i,t)$ and $C^{(2)}(q_i,t)$ begin close to 0 and 1, respectively, as expected, and as the coupling kicks in and the atom decays, one might expect them to evolve to 1 and 0, respectively. This is in fact correct for almost all $q_i$, however non-uniformly in $q_i$. As expected from the nature of the bilinear coupling Hamiltonian Eq.~(\ref{Hmodel}), the conditional electronic amplitude associated with larger photonic displacements $q_i$ couple more strongly than those associated with smaller ones, so the conditional amplitude on the upper surface falls away from 1 starting on the outer edges and then moving in. In fact, the conditional amplitude at $q=0$ remains forever stubbornly at the upper surface, unaffected by the coupling to the field. 

This non-uniformity is reflected in the q-TDPES, and leads to a strong deviation from the harmonic form it has in the absence of matter. The potential, driving the photonic motion, loses its harmonic form in the initial time steps as the decay begins, peeling away starting from the outer $q_i$.  The potential nearer $q_i = 0$ remains harmonic for the initial stages, but as time goes on, more of the surface peels away from the upper surface, while a  peak structure develops near $q_i=0$ that gets increasingly localized and increasingly  sharp as the atom decay process completes and the photon is fully emitted. It is this peak structure in the potential driving the photonic system that excites the system from the zero-photon state towards the one-photon state.

We turn now to the components of this exact surface. In the weighted BO surface, $\epsilon_{\rm wBO}(q_i,t)$ that is plotted in the middle right panel, we see the same peeling away from the outer edges, but sticking resolutely to the original upper surface at $q_i = 0$. As the decay occurs, $\epsilon_{\rm wBO}(q_i,t)$  gradually melts to the lower surface everywhere except for a shrinking region near the origin that sticks to the upper surface. The peak seen in the full q-TDPES on the other hand comes from $\epsilon_{\rm kin}(q_i,t)$, plotted in the lower left panel, which gets sharper and sharper as the photon is emitted. Mathematically, this structure follows from the change in the conditional-dependence of $\Phi_\dulq$ near $q_i=0$,  as the electronic state associated with $q_i=0$ remains on the upper qBO surface while away from $q=0$, in a shrinking region, the electronic state is associated with the lower surface. This gets sharper as $\chi(q_i=0,t)$ gets smaller and smaller there. 
%A spike structure was also observed in the electron-nuclear factorization within a two-level system at nodes of the marginal density~\cite{L15}. 
One can show from the analytic solution, that, in the long-time limit, the surface at $q_i = 0$ grows exponentially with $t$ at a rate determined by $\Gamma$, while for $q\neq 0$, $\epsilon_{\rm kin}(q_i\neq0, t\to\infty) \to 0$. 

These features of $\epsilon_{\rm wBO}$ and $\epsilon_{\rm kin}$ are very similar for both the cross-section that cuts along the resonant mode (Fig.~\ref{fig:WW-on}) and that cutting along the slightly off-resonant (Fig.~\ref{fig:WW-off}). 
The remaining component of the q-TDPES, $\epsilon_{\rm GD}$ is much smaller than the other components, and has a different structure in the two cases. In fact, it is straightforward to show from the analytic solution that $\epsilon_{\rm GD}(q_i = 0,t)$ is independent of $t$, and that uniformly shifting \linebreak $\epsilon_{\rm GD}(q_i,t)$ so that  $\epsilon_{\rm GD}(q_i = 0,t) \equiv 0$ yields $\epsilon_{\rm GD}(q_i \neq 0, t \gg \Gamma) \to \omega_0 -\omega_i$ for $q_i$ large. That is, there is a symmetric step-like feature in $\epsilon_{\rm GD}$, of the size of the difference in the mode frequency of interest and the resonant mode, and as $t$ gets larger, this feature sharpens.

Thus, we can see that in the Wigner-Weisskopf limit, the potential driving the photonic modes deviates significantly from its initial harmonic form during the decay, although once again becoming harmonic almost everywhere (except at $q=0$) in the long-time limit. The atom-photon correlation is required to capture these effects, and if one wanted to model this exact q-TDPES, the conditional dependence of the electronic amplitude is crucial to include.

\subsection{Two-Level System Coupled to a Single Resonant Mode}
\label{sec:tls}
We now turn to the other limit, and tune the cavity so that there is just one mode that couples appreciably to the two-level atom, with a mode frequency that is resonant with the atomic energy difference. 

The q-BO surfaces can be easily found by diagonalizing $H_{\rm qBO}$ of Eq.~(\ref{HqBO}), keeping only one mode with $\omega_\alpha = \omega_0$ in the field:
\ben
\epsilon_{\rm qBO}(q) = \frac{1}{2}\omega_0^2 q^2 \mp \sqrt{\omega_0^2/4 +(d_{eg}\lambda\omega_0)^2q^2}
\een
For couplings $\lambda d_{eg}\ll 1/2$, the q-BO surfaces are approximately parallel and harmonic except at large $q$ (see also Ref.~\cite{Flick2017a}). So in this case if the initial photonic state is a vacuum, then the ensuing dynamics is driven by a largely harmonic potential, without much perturbation from the atom, except at larger $q$. Deviations from parallel harmonic surfaces, and hence non-qBO behavior, occurs at larger $q$ and as the coupling increases. We will investigate  the exact q-TDPES driving the photonic dynamics for three different coupling strengths, (dl = $\{0.01,0.1,0.4\}$) and will include a plot of the two qBO surfaces with our results for comparison with the exact q-TDPES.

\begin{figure}
 \centering
\includegraphics*[width=1.0\columnwidth]{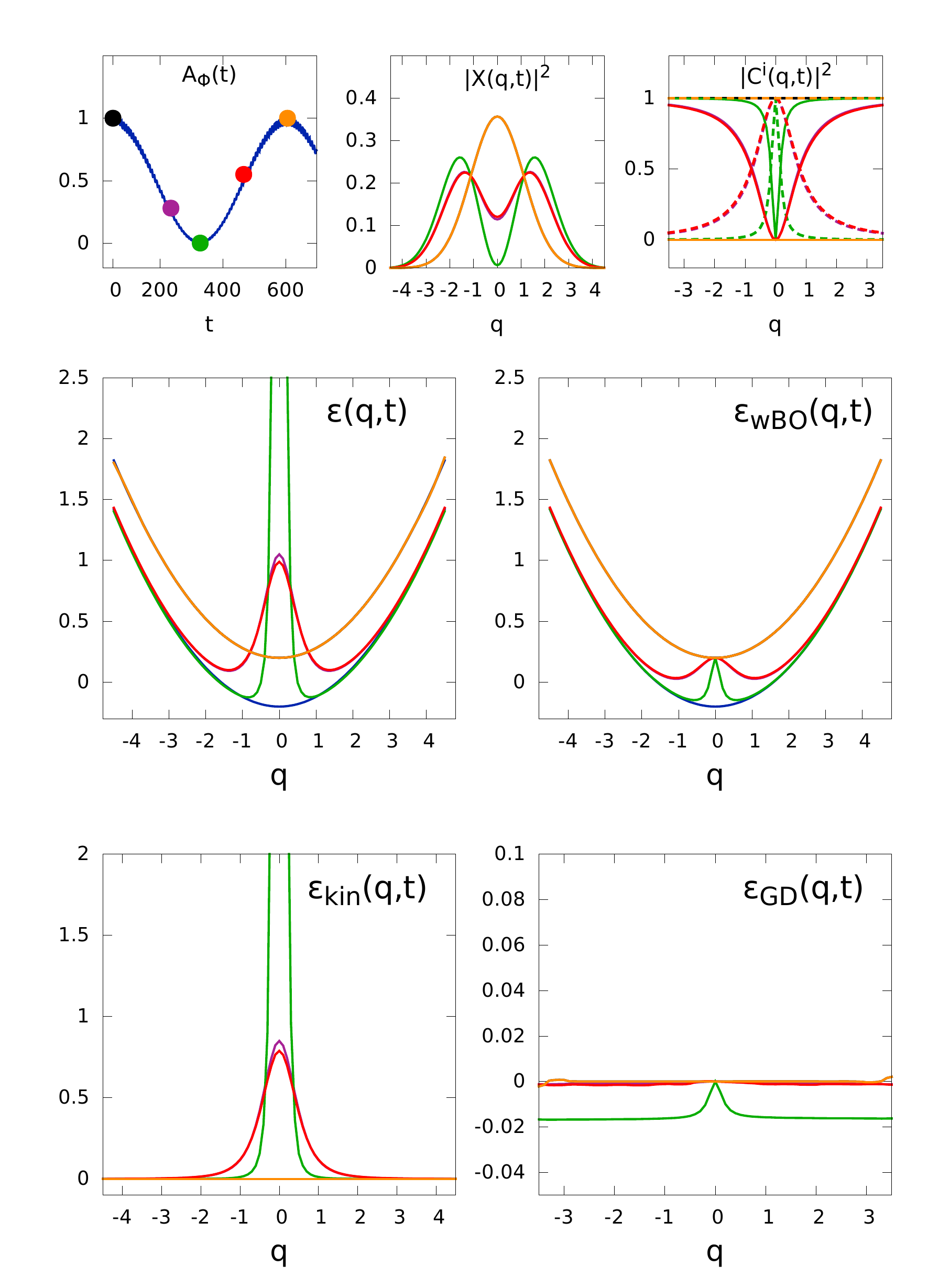}
 \caption{The $\epsilon(q,t)$ for the excited BO initial state and coupling strength $dl=0.01$. The top left panel shows $A_{\Phi}(t)$, where the different colored dots depict the times of the dynamics within this plot. The top middle and right plots show the photonic distribution $\vert\chi(q,t)\vert^2$ and the electronic coefficients in the BO basis,  $|C^{(1)}(q_i;t)|^2$ (dashed), $|C^{(2)}(q_i;t)|^2 $ (solid), for the time snapshots shown. The middle and lower panel show the  q-TDPES $\epsilon(q,t)$ and its decomposition into components for the given time snapshots. The q-BO surfaces are shown in the middle panel in blue for reference.}
  \label{fig:coupled_0.01}
\end{figure}

\begin{figure}
 \centering
\includegraphics*[width=1.0\columnwidth]{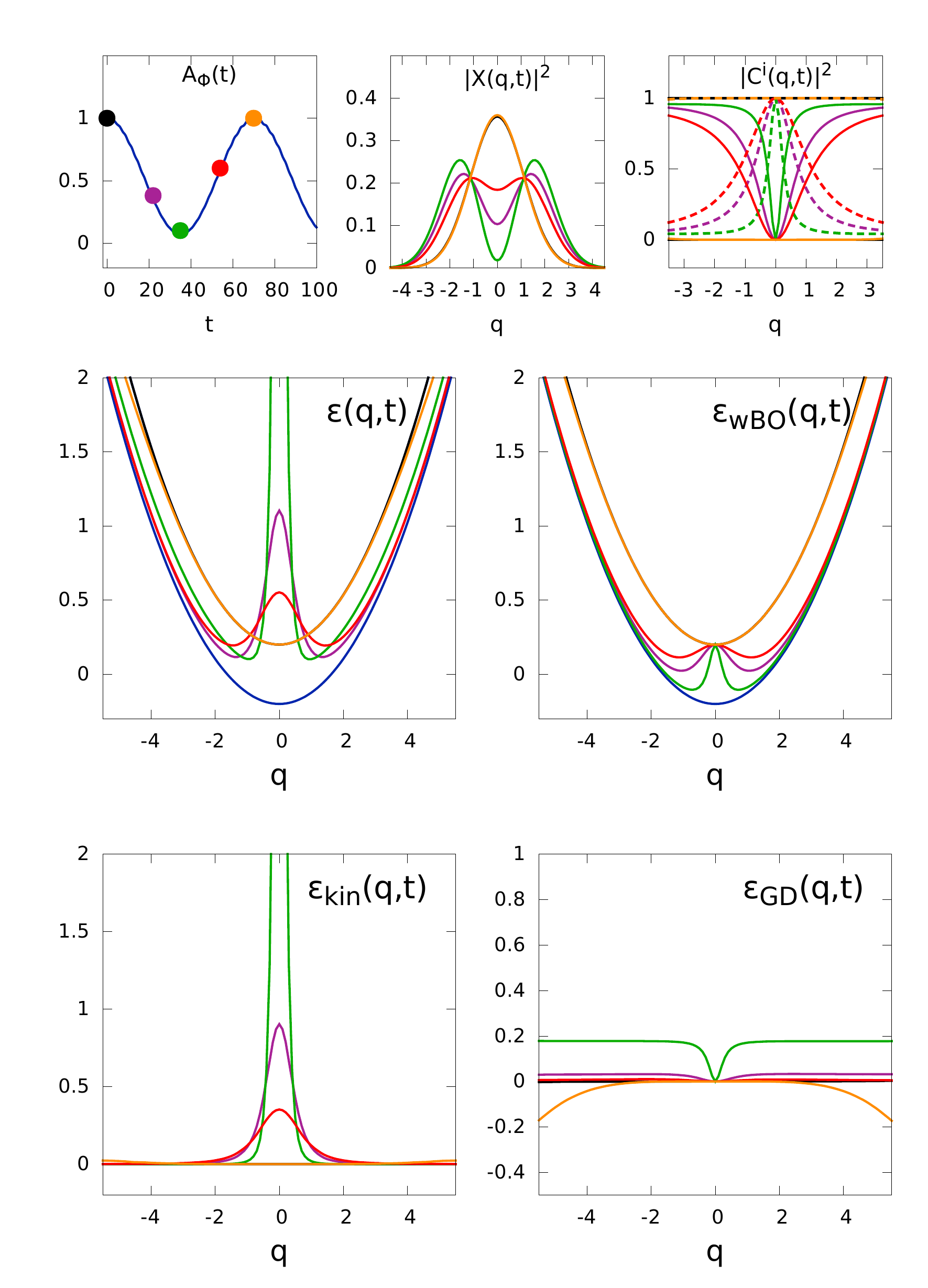}
 \caption{As  in Figure~\ref{fig:coupled_0.01} but with coupling strength $d_{eg}\lambda = 0.1$.}
  \label{fig:coupled_0.1}
\end{figure}

\begin{figure}
 \centering
\includegraphics*[width=1.0\columnwidth]{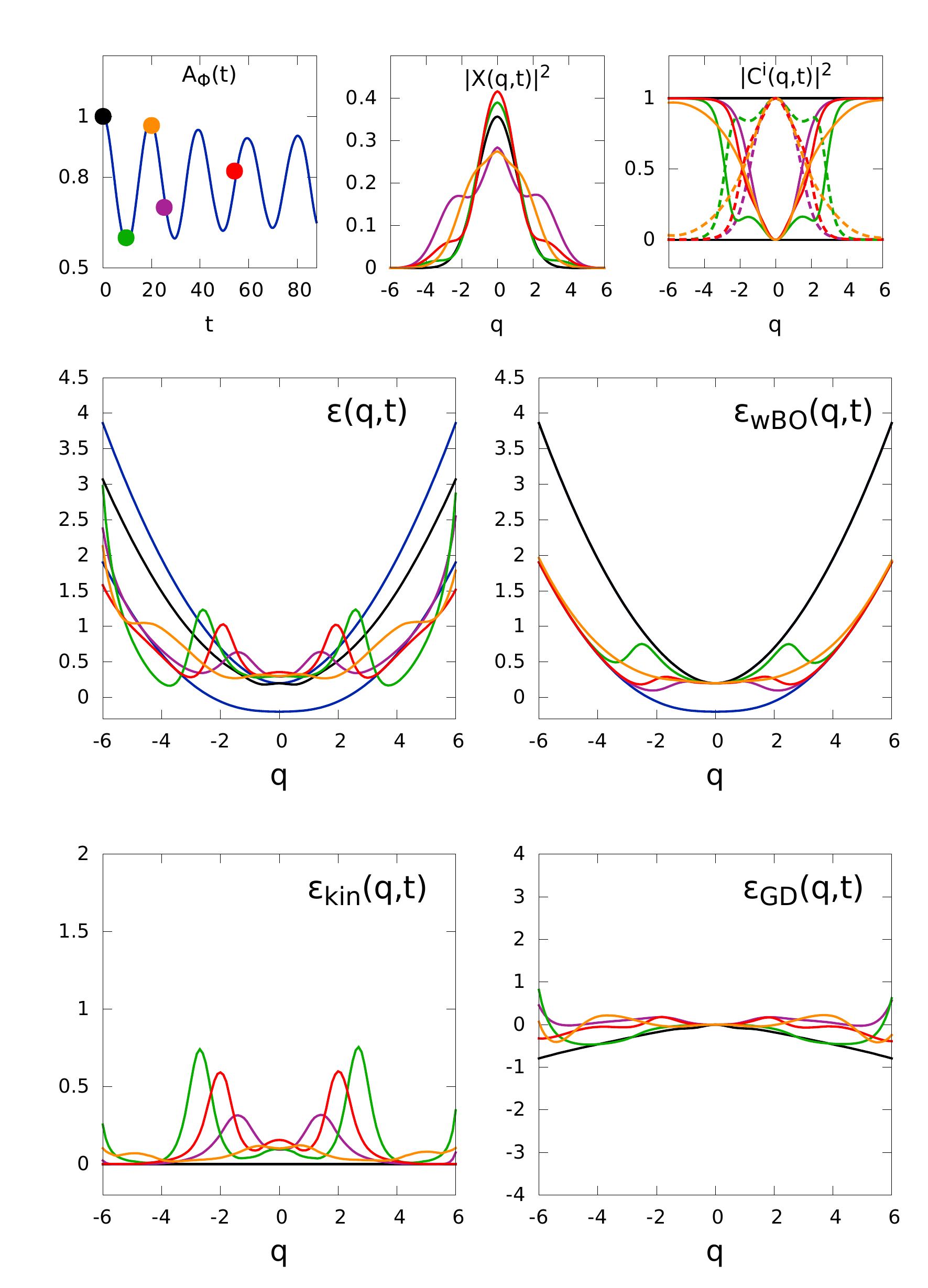}
 \caption{As  in Figure~\ref{fig:coupled_0.01} but with coupling strength $d_{eg}\lambda = 0.4$. }
  \label{fig:coupled_0.4}
\end{figure}

\begin{figure}
 \centering
\includegraphics*[width=1.0\columnwidth]{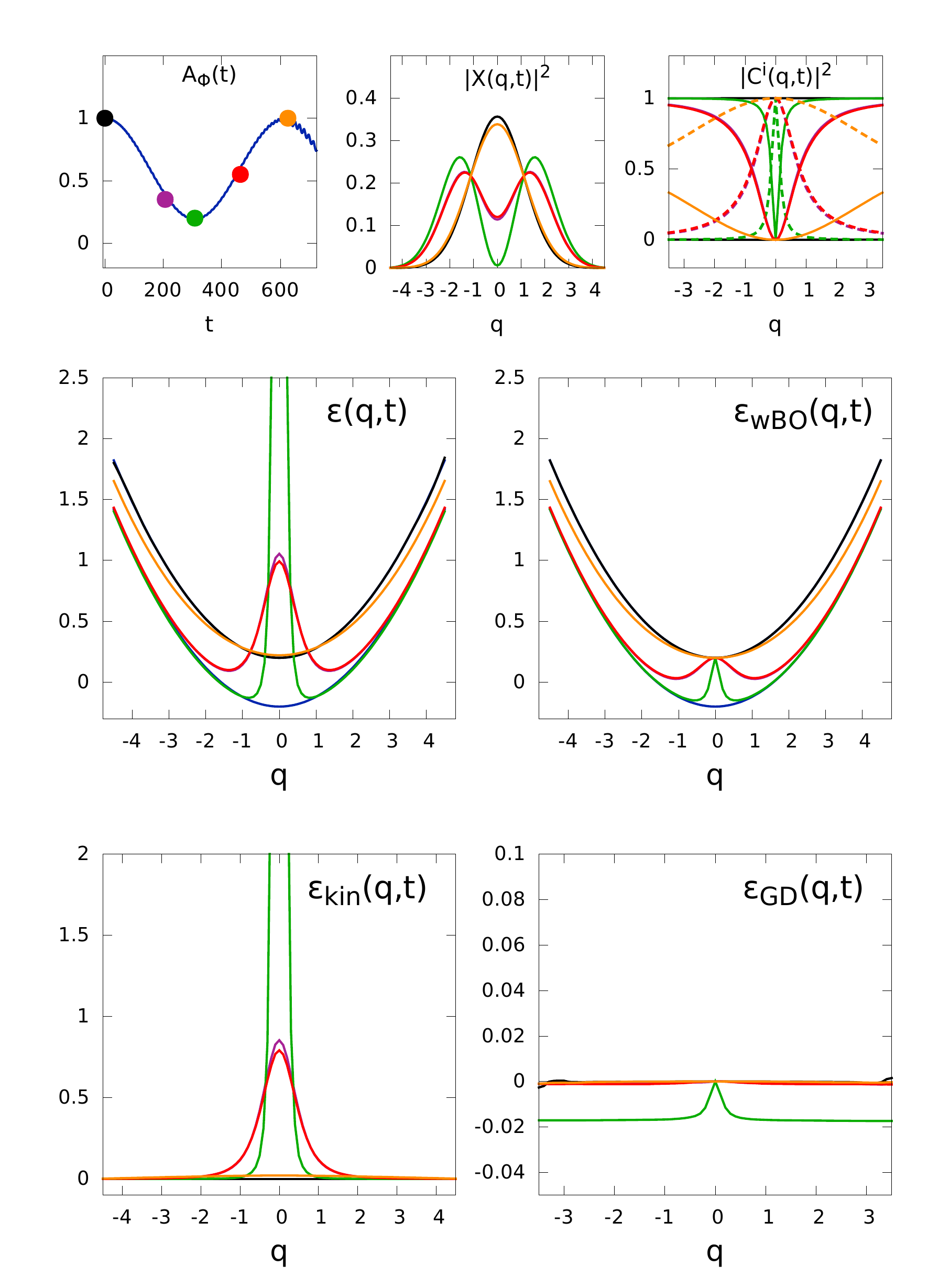}
 \caption{As  in Figure~\ref{fig:coupled_0.01}  with coupling strength $d_{eg}\lambda = 0.01$ but with the initial purely factorized state. }
 \label{fig:uncoupled_0.01}
\end{figure}

In Figure~\ref{fig:coupled_0.01} we plot the exact q-TDPES for coupling strength $d_{eg}\lambda = 0.01$ beginning with the atom in the excited BO level, multiplied by the photonic ground-state. 
On the upper panel (left) we plot the autocorrelation function 
\begin{equation}
A_\Psi(t) = \left\vert\int dq  \orr{\Psi}^\dagger(q,0)\cdot\orr{\Psi}(q,t)\right\vert^2
\end{equation}
to indicate the approximate periodicity of the system dynamics. 
Comparing with Figures~\ref{fig:coupled_0.1} and~\ref{fig:coupled_0.4},  we find a decrease of the approximate period with the increase of coupling strength until the periodicity breaks down for the strong coupling $dl = 0.4$. Further, we observe from Fig~\ref{fig:uncoupled_0.01} that the periodicity depends on the choice of initial state, as the time for one period decreases if we choose the initial state to be fully factorized. The weakest coupling strength we have chosen is on the borderline of being in the Rabi regime~\cite{YE85}, while the strongest is far from it.

\begin{figure}
 \centering
\includegraphics*[width=1.0\columnwidth]{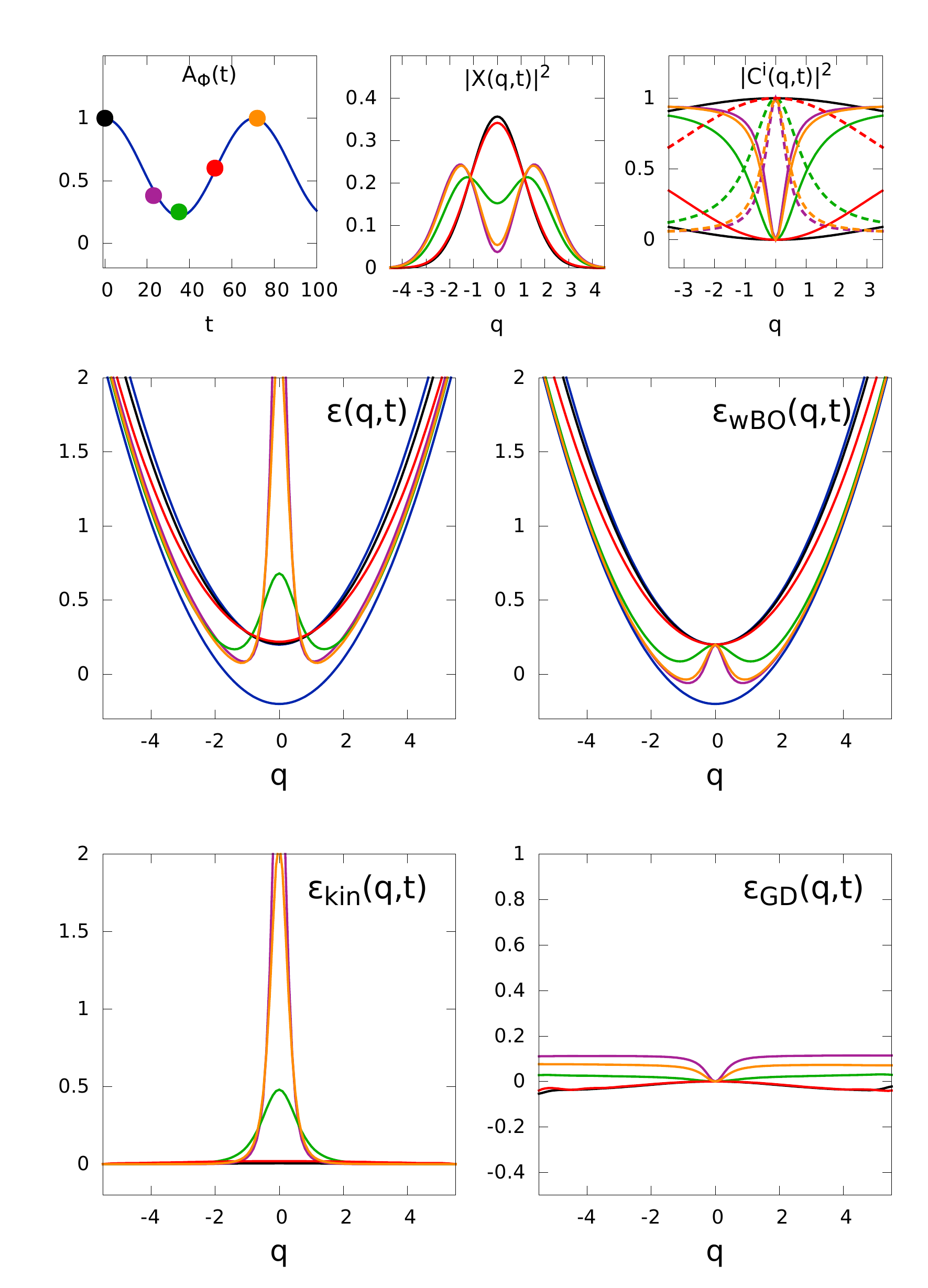}
 \caption{As  in Figure~\ref{fig:coupled_0.1} with coupling strength $d_{eg}\lambda = 0.1$ but with the initial purely factorized state.}
  \label{fig:uncoupled_0.1}
\end{figure}

\begin{figure}
 \centering
\includegraphics*[width=1.0\columnwidth]{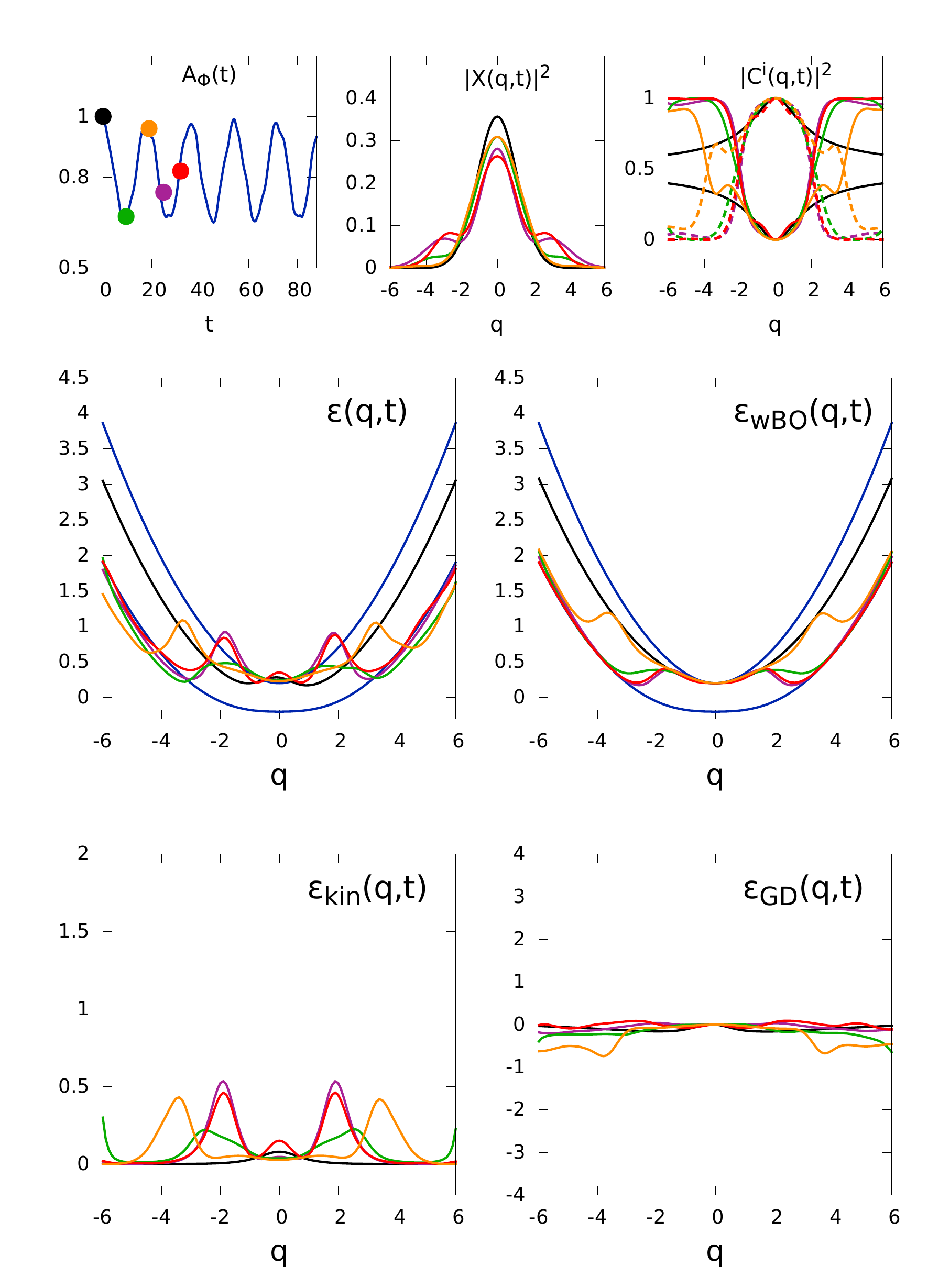}
 \caption{As  in Figure~\ref{fig:coupled_0.4} with coupling strength $d_{eg}\lambda = 0.4$ but with the initial purely factorized state.}
  \label{fig:uncoupled_0.4}
\end{figure}

The photonic distribution (middle) and conditional electronic coefficients (right) are shown in the topmost panel of  Figs.~(\ref{fig:coupled_0.01}) - (\ref{fig:uncoupled_0.4}). The photonic system begins in the vacuum state, while the electronic distribution starts, as defined in the initial condition, with an electron in the excited state and no electron in the ground-state, either BO or field-free. The initial coefficients are $C^{(1)}(q,0) =1$ and $C^{(2)}(q,0) =0$ when beginning in the BO states. When beginning in the fully factorized  state, these coefficients deviate from these uniform values, especially for larger $q$, with deviation increasing with the coupling strength.

The dynamics depends significantly on whether the initial state is the correlated qBO state  (Figs. 5--7) or a fully factorized one (Figs. 8--10). The fully factorized initial state would be  the physical one when an excited atom is  instantaneously brought into a closed cavity and  just then its dynamics is studied, while the excited qBO state results when there is initially an external dissipative coupling together with an applied  resonant field to maintain the atom in that excited state before the dynamics is examined. We turn now to the dynamics and to the structure of the exact q-TDPES for this latter case first.

 After some time we see a transition of the electron from the excited state to the ground-state as indicated by these coefficients. We observe that the transfer begins earlier for higher values of $q$ and then is followed by lower $q$-values, but again the conditional amplitude at $q=0$ sticks to the upper surface at all times in all cases as there is no coupling for $q=0$. The $q$-dependence of these coefficients has a significant role in shaping the structure of the q-TDPES that we will shortly discuss. At the same time, the probability of photon emission increases, as indicated by the morphing of the initial gaussian in $\chi(q,t)$ towards its first-excited profile. For the weakest coupling strength after a half period, the system begins to move back approximately to its initial state, as the photon is reabsorbed and atom becomes excited again. For strong coupling $dl=0.4$ the periodic character is lost and we find more wells and structure appearing in the \linebreak displacement-field density profile. With such strong coupling the qBO surfaces are quite distorted from a pure harmonic, as evident in the plot (blue lines in the middle panel), and the anharmonicity brings more frequencies into play. A one-photon state that is associated with the lower q-BO surface has a wider profile with density maxima further out than a one-photon state associated with the upper surface would have, for example. In fact the character of the coupled cavity-matter system becomes quite mixed, as is evident from the conditional electronic coefficients shown on the right, and as one goes along the photonic coordinate $q$ one associates with different superpositions of the electronic states. This leads to interesting structure in the exact q-TDPES, that, when decomposed in terms of the q-BO surfaces, has components that vary a lot with $q$ (i.e. not just a piecewise combination).

The q-TDPES for initial state prepared in the upper q-BO state begins with the weighted BO component, $\epsilon_{\rm wBO}$ on top of the upper q-BO surface as expected. 
For the weakest coupling, $d_{eg}\lambda = 0.01$, $\epsilon_{\rm wBO}(q,t)$ then melts down to the lower surface over half a cycle, peeling away from  the outer higher $q$-values first, in a similar way to what was seen in the Wigner-Weisskopf limit. This potential approaches the lower surface before returning back to the upper BO-surface, but the region near $q=0$ remains bound to the upper surface. The time-dependent double-well structure in the potential is again important in driving the photon emission. A similar trend is seen for the stronger coupling 0.1 in Fig.~\ref{fig:coupled_0.1}, but for the strongest coupling  $d_{eg}\lambda=0.4$, $\epsilon_{\rm wBO}(q,t)$ shows a more complicated correlation in $q$, with structures mirroring those in the displacement-field density discussed above.
As for the kinetic component, for the weaker couplings,  a peak structure in $\epsilon_{\rm kin}(q,t)$ develops that grows and narrows during the photon emission stage, similar to what was seen in Wigner-Weisskopf, but  this then reverses during the reabsorption here. Again for the stronger coupling, the structure is more complicated, mirroring the more complicated dynamics. 
The gauge-dependent part, $\epsilon_{\rm GD}$ is generally a smaller contribution to the total q-TDPES compared to the other components, but again we see step-like features for the weaker couplings, and more complicated dynamics for the strongest coupling.

For the fully factorized initial states, although  the photonic field still begins in the vacuum state, the electronic state is not purely in the upper BO surface; the electronic state associated with larger $q$ has already some component in the ground-state. So at these larger values of $q$, the initial $\epsilon_{\rm wBO}(q,0)$ surface dominates the q-TDPES and is anharmonic from the very start, lying intermediate between the upper and lower q-BO surface. In the weak coupling case, the differences are only large at values of $q$ much larger than shown in the plot, and these are physically unimportant given there is very little photonic field probability there; hence Fig. 5 and 8 are almost identical. For strong couplings, comparing Fig 7 and 10 shows that the q-TDPES has a tamer structure for the fully-factorized initial state than for the correlated q-BO initial state, especially at larger $q$; this is likely because less energy is available at these larger $q$ for the system to exchange between the atomic and photonic systems because the atomic state correlated with large $q$  is not completely in its excited state initially. 

To summarize:  At time zero the exact q-TDPES starts on the upper q-BO-surface, which, depending on coupling strength and choice of initial state, ranges from lying directly on top of the upper q-BO surface (weaker coupling and with q-BO initial state), to in between the two q-BO surfaces with deviations from the upper being larger for larger $q$ (stronger coupling, or fully factorized initial state).  After some time the potential starts to melt down onto the lower BO-surface, first starting at higher $q$-values and then followed by lower $q$-values, with peak structures developing in the interior region. Around $q = 0$ the kinetic-component dominates, which leads to an increasing and after half a period decreasing peak. For stronger coupling we observe several peak features in the potential and significant deviations from the curvature of the q-BO surfaces throughout $q$ small contribution below the lower BO-surface; the deviations at larger $q$ arise from the gauge-dependent component.

\section{Summary and Outlook}
\label{sec:summary}
We have introduced here an extension of the exact- factorization approach, that was originally derived for coupled electron-nuclear systems, to light-matter systems in the non-relativistic limit within the dipole approximation. We have presented different possible choices for the factorization but in this work have focussed on the one where the marginal is chosen as the photonic system and the matter system is then conditionally-dependent on this. This choice is particularly relevant when one is primarily interested in the the state of the radiation field since the exact factorization yields a time-dependent Schr\"odinger equation for the marginal, while the conditional is described by an equation with an unusual matter-photon coupling operator. The equation for the marginal is, in a sense, simpler than that in the electron-nuclear case, since the vector potential, q-TDVP,  appearing in the equation can always be gauged away into a scalar potential, the q-TDPES. We have studied the potential appearing in this equation in a gauge where the q-TDVP is zero, for a two-level system coupled to an infinite number of modes in the Wigner-Weisskopf approximation, and for a two-level system coupled to a single photonic field mode with a range of coupling strengths. In all cases we find a very interesting structure of the potential that drives the photonic dynamics, and in particular, large deviations from the harmonic form of the free-photon field. These deviations completely incorporate the effect of the matter system on the photonic dynamics. We also studied the effect of beginning in an initially purely factorized light-matter state, compared to a q-BO initial state, finding significant differences for larger coupling strengths in the ensuing dynamics, implying that in modelling these problems a careful consideration of the initial state is needed.

To use the exact factorization for realistic light-matter systems, approximations will be needed, since solving the exact factorization equations is at least as computationally expensive as solving the Schr\"odinger equation for the fully coupled system. The success of such an approximation depends on how well the q-TDPES is modelled. The components of the exact q-TDPES beyond the weighted BO depend significantly on the $q$-dependence of the conditional probability amplitude; approximations that neglect this dependence (Ehrenfest-like) will likely lead to errors in the dynamics. 
It has been shown recently that mixed quantum-classical trajectory methods that are derived from the exact factorization approach can
correctly capture decoherence effects \cite{doi:10.1021/acs.jpclett.7b01249,AMAG16,MAG15}.
Since photons are intrinsically non-interacting and therefore even simpler to treat than nuclei, we expect in analogy to the electron-nuclear case
that semiclassical trajectory methods
derived from systematic and controlled approximations to the full exact factorization of the light-matter wavefunction will be able to capture
decoherence effects beyond the Ehrenfest limit for light-matter coupling. This will be subject of future investigations.

\begin{acknowledgement}
We acknowledge financial support from the European Research Council(ERC-2015-AdG-694097), Grupos Consolidados (IT578-13), and  European Union's H2020  programme under GA \linebreak no.676580 (NOMAD) (NMH, HA, and AR).
Financial support from the US National Science Foundation
CHE-1566197
is also gratefully acknowledged (NTM).

\end{acknowledgement}

\bibliographystyle{epj}
% \bibliography{}
\bibliography{./qed}

\end{document}